\documentclass[12pt]{article}

\usepackage{a4}
\usepackage{a4wide}

\usepackage{amsmath}
\usepackage{amscd}
\usepackage{amsfonts}
\usepackage{amssymb}
\usepackage{mathrsfs}
\usepackage{fancybox} 
\usepackage{enumerate}

\usepackage{bm}
\usepackage{amscd}
\usepackage[dvips]{color}
\usepackage{graphicx}
\usepackage{subfigure}

\makeatletter

\@addtoreset{equation}{section}
\makeatother

\usepackage{amsthm} % This is for \newtheorem*
\newtheorem*{pf}{Proof}
\def\tr{\mathrm{tr}}
\def\Tr{\mathrm{Tr}}

\def\half{{1\over2}}
\def\nn{\nonumber\\}
\def\sgn{\mathrm{sgn}}

\newcommand{\wt}{\widetilde}

\def\={\stackrel{\bullet}{=}}

\def\({\left(}
\def\){\right)}
\def\[{\left[}
\def\]{\right]}

\def\cL{{\cal L}}

\def\cN{{\cal N}}
\def\cO{{\cal O}}

\def\mbf{\mathbf}

\def\mf {\mathfrak }

\def \be {\begin{equation}}
\def \ee {\end{equation}}
\def \bea {\begin{eqnarray}}
\def \eea {\end{eqnarray}}
\def \beal#1 {\begin{align}#1\end{align}}
\def \bes#1 {\begin{equation}\begin{split}#1\end{split}\end{equation}}
\def \nn {\notag\\}

\def\bra#1{\langle #1 |}
\def\ket#1{|#1 \rangle}
\def\aver#1{\left\langle #1 \right\rangle}

\makeatletter

\@addtoreset{equation}{section}
\makeatother

\begin{document}

\begin{titlepage}
% \vspace{-2cm}
\title{
\begin{flushright}
\normalsize{ 
YITP-16-49\\
April 2016}
\end{flushright}
       \vspace{1.5cm}
Scattering Amplitude and Bosonization Duality in General Chern-Simons Vector Models
\vspace{1.5cm}
}
\author{
Shuichi Yokoyama\thanks{E-mail: shuichi.yokoyama[at]yukawa.kyoto-u.ac.jp
}
\\[25pt] 
{\it Yukawa Institute for Theoretical Physics, Kyoto University,}\\
{\it Kyoto 606-8502, Japan}
\\[10pt]
{\it Department of Physics,  Keio University,}\\
{\it Research and Education Center for Natural Sciences,}\\
{\it Kanagawa 223-8521, Japan}
\\[10pt]
% {\small \tt E-mail: shuichitoissho(at)gmail.com}
}

\date{}
\maketitle

\thispagestyle{empty}

% \vspace{.2cm}

\begin{abstract}
\vspace{0.3cm}
\normalsize

We present exact large $N$ calculus of four point function in general Chern-Simons bosonic and fermionic vector models. 
Applying the LSZ formula to the four point function we determine two body scattering amplitudes in these theories taking a special care for a non-analytic term to achieve unitarity in the singlet channel. 
We show that the S-matrix enjoys the bosonization duality, unusual crossing relation and non-relativistic reduction to Aharonov-Bohm scattering. We also argue that the S-matrix develops a pole in a certain range of coupling constants, which disappears in the range where the theory reduces to Chern-Simons theory interacting with free fermions.    

\end{abstract}
\end{titlepage}
%\tableofcontents

%%%%%%%%%%%%%%%%%%%%%%%%%%%%%%%%%%%%
\section{Introduction}
\label{intro}

Three dimensional vector model gauged by Chern-Simons interaction (Chern-Simons vector model) has turned out to be an intersection of various subjects in theoretical physics. 
An initiative study was done in \cite{Giombi:2011kc,Aharony:2011jz}, where it was conjectured that Chern-Simons vector model inherits large $N$ solubility of vector model under the 't Hooft large $N$ limit and becomes conformal field theory (CFT) dual to parity-violated higher-spin gravity theory in AdS$_4$ \cite{Maldacena:1997re} by extending the conjecture of vector model/higher-spin theory \cite{Klebanov:2002ja}. (See \cite{Chang:2012kt} for its supersymmetric extension.)
 
Large $N$ exact calculus of three point correlation functions of higher spin currents was performed \cite{Aharony:2012nh,GurAri:2012is}
in accordance with general study of correlators including parity non-invariant contact terms of the class of three dimensional CFT \cite{Giombi:2011rz,Maldacena:2011jn,Maldacena:2012sf}. (See also \cite{Nizami:2013tpa,Bedhotiya:2015uga}.) 
Analysis of correlation functions indicated that there exists a bosonization duality between a pair of bosonic Chern-Simons vector model and fermionic one as a generalization of level-rank duality known in pure Chern-Simons theory by including vector matter fields. This novel duality was confirmed from the large $N$ thermal free energy of Chern-Simons vector models \cite{Aharony:2012ns,Jain:2013py,Jain:2013gza,Yokoyama:2013pxa,Minwalla:2015sca}.
(See also \cite{Giombi:2011kc,Jain:2012qi,Yokoyama:2012fa,Takimi:2013zca,Moshe:2014bja}.)
Evidence for this duality to hold during renormalization group (RG) flow was provided in \cite{Jain:2013gza,Minwalla:2015sca,Gur-Ari:2015pca}, which  indicated that this bosonization duality in three dimensions is obtained by continuous deformation from supersymmetric duality known as Seiberg-like duality \cite{Giveon:2008zn,Benini:2011mf}. (See also \cite{Banerjee:2012gh,Frishman:2013dvg,Bardeen:2014paa,Gurucharan:2014cva,Radicevic:2015yla} for study of other aspects.)

Three dimensional bosonization duality was uncovered from large $N$ exact results of scattering amplitudes at the level of elementary particles \cite{Jain:2014nza,Inbasekar:2015tsa}. (See also \cite{Geracie:2015drf}.) 
It turned out that the scattering amplitude in S-channel computed from Euclidean theory via Wick rotation becomes non-unitary \cite{Jain:2014nza}, which was also observed in the original Aharonov-Bohm scattering.  
The non-unitarity problem in the non-relativistic case was resolved by carefully taking into account the singular contribution in the forward scattering \cite{Ruijsenaars:1981fp}. 
The resolution of non-unitarity puzzle in the relativistic case was similarly developed by taking care of the forward scattering so as to restore unitarity \cite{Jain:2014nza}.
Then it was shown that the improved S-matrix exhibits bosonization duality, which just exchanges bosonic particles in one theory with fermionic ones in the dual one. 
By combining this with the fact that a bosonic or fermionic particle interacting with Chern-Simons gauge field acquires some statistics to be anyon \cite{Zhang:1988wy,Fradkin:1991wy}, the bosonization in three dimensions can be naturally understood as duality between two theories of anyon. 

Along the line this paper aims at performing the large $N$ exact calculus on scattering amplitudes and confirming the duality therefrom in a {\it general} Chern-Simons vector model, which includes the double and triple trace terms in bosonic Chern-Simons vector model or quadratic and cubic ones of the auxiliary field in the fermion one.
The bosonization duality is expected to hold for the general Chern-Simons vector models since these models connect the regular Chern-Simons vector models and the critical ones by RG flow. Indeed the duality in the general case was confirmed from exact large $N$ calculation of the thermal free energy \cite{Minwalla:2015sca}. 
The calculus of S-matrix in the bosonic side is straightforward  by adding the contribution of the triple trace coupling into that in  \cite{Jain:2014nza}, while that in the fermionic side is rather non-trivial since interaction of the auxiliary field generates ultra-violet (UV) divergent integrals as fermion loops in Gross-Neveu model \cite{Gross:1974jv,Rosenstein:1988pt}, which are non-renormalizable in the sense of weak coupling expansion and have to be regularized suitably in $1/N$ expansion. 

The rest of this paper is organized as follows. 
In the section \ref{CSFVM} we set up the system and perform some preliminary large $N$ analysis.  
In the section \ref{4pt} we compute the four point function of the fermionic field exactly in the light-cone gauge with a specific momentum frame.
By applying the LSZ formula to the four point function we determine the S-matrix in T-channel in Section \ref{Tchannel}, and that in S-channels with taking account of non-analytic contribution in the forward scattering in Section \ref{Smatrix}. Section \ref{discussion} is devoted to conclusion and discussion. In Appendix we detail regularization skipped in the main text (\ref{regularization}), construct asymptotic states of scattering process (\ref{states}), and give a brief derivation of scattering amplitudes in general Chern-Simons bosonic vector model (\ref{bosonicSmatrix}).

%%%%%%%%%%%%%%%%%%%%%%%%%%%%%%%%%%%%
\section{General Chern-Simons fermion vector model}
\label{CSFVM}

\begin{figure}[th]
  \begin{center}
\includegraphics[scale=.5]{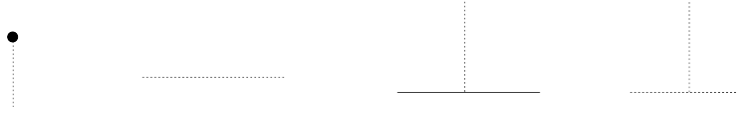}
\caption{Feynman rule involving the auxiliary field $\sigma_F$ is depicted. The solid and dotted lines represent the fermion and the auxiliary field, respectively. From the left, the diagrams describe tadpole, tree-level {\it propagator}, Yukawa interaction with fermion, and cubic self-interaction.  }
\label{vertex}
  \end{center}
\end{figure}

In this section we perform preliminary analysis on general $U(N)_{k_F}$ Chern-Simons fermion vector model in the 't Hooft large $N$ limit, $N, k_F \to \infty$ with $\lambda:=N/k_F$ fixed.%
\footnote{ 
We basically use the same convention adopted in \cite{Minwalla:2015sca}. Under the convention the 't Hooft coupling constant is bounded so that $|\lambda|\leq1$.
Especially upon encountering divergent integrals we regularize them by dimensional regularization in the way described in \cite{Jain:2012qi}.  
}
Lagrangian of general $U(N)_{k_F}$ Chern-Simons fermion vector model is defined by using an auxiliary field $\sigma_F$ as 
\bes{
~\label{csfnonlinear2}
\cL_F =& i \varepsilon^{\mu\nu\rho} {k_F \over 4 \pi}
\Tr( A_\mu\partial_\nu A_\rho -{2 i\over3}  A_\mu A_\nu A_\rho)
+  \bar{\psi} \gamma_\mu D^{\mu} \psi \\
&~~~~+\sigma_F (\bar\psi \psi - {k_F \over 4 \pi} y_2^2) - {k_F \over 4 \pi} y_4 \sigma_F^2 + {k_F \over 4 \pi} y_6 \sigma_F^3 
}
where 
\be
D_{\mu} \psi =(\partial _{\mu} -i A_\mu ) \psi \quad 
D_{\mu} \bar \psi =(\partial _{\mu} \bar \psi + i  \bar \psi A_\mu ).
\ee
Here we abbreviate the gauge, spinor and space indices. 
For convenience we depict Feynman rule involving the auxiliary field in Figure \ref{vertex}.
This theory is not renormalizable in the weak coupling expansion whereas it is in $1/N$ expansion. In particular when the cubic self-interaction of the auxiliary field vanishes this system is the same as three dimensional Gross-Neveu model \cite{Gross:1974jv} gauged with Chern-Simons interaction. In contrast to the fact that two dimensional Gross-Neveu model is asymptotically free, three dimensional one has the ultra-violet fixed point \cite{Rosenstein:1988pt}.
In the current setup the critical theory is obtained by a limit such that $y_2^2, y_4, y_6 \to 0$.
Note that the auxiliary field has odd parity since the Gross-Neveu model is classically parity invariant.  

The exact propagator of the fermionic field in the large $N$ limit was already computed in \cite{Minwalla:2015sca}. The result reads 
\beal{
\aver{\psi^m(p) \bar\psi_n(-p')} = \delta^m_n(2\pi)^3\delta^3(p-p')\alpha_F(p), \quad 
\alpha_F(p) = {1 \over i\gamma^\mu p_\mu + \Sigma_F(p) }.
\label{fermionpropagator}
} 
Here $-\Sigma_F(p)$ is 1PI self energy of fermion given by
\bes{
\Sigma_F(p) =& \Sigma_I(p) + \Sigma_+(p) \gamma^+, \\
\Sigma_I(p) =& \lambda \sqrt{p_s^2+c_F^2} + (\sgn(\lambda) -\lambda)c_F, \\
\Sigma_+(p) =& ip_+ {c_F ^2 -\Sigma_I(p)^2 \over p_s^2}, 
\label{1piselfenergy}
}
where $c_F$ is the physical mass of fermion determined by the following gap equation
\be 
(1-3y_6)  (\sgn(\lambda)-\lambda)^2 c_F^2 + 2y_4 (\sgn(\lambda)-\lambda) c_F + y_2^2 -c_F^2 = 0.
\label{cf}
\ee
Note that we consider a situation where $c_F>0$. 

\begin{figure}[thbp] 
  \begin{center}
  \includegraphics[scale=.5]{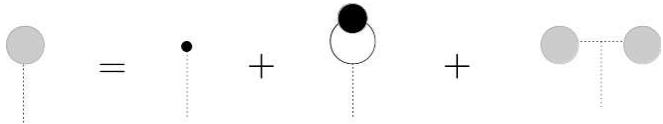}  
  \end{center}
  \vspace{-0.5cm}
  \caption{The bootstrap diagram for the exact tadpole is drawn. Here the black node represents the exact propagator of fermion and the gray node describes the exact tadpole diagram. }
\label{tadpolebootstrap}
\end{figure}

For later convenience we determine the exact tadpole diagram contribution.
The tadpole diagram satisfies the bootstrap diagram depicted in Figure \ref{tadpolebootstrap}, which reads
\be 
T = {k_F \over 4\pi}y_2^2 -N \int {d^3p\over (2\pi)^3} (-\tr {1 \over i\gamma^\mu p_\mu + \Sigma_F(p) }) -3{k_F \over 4\pi}y_6 \times ({1\over-2{k_F \over 4\pi} y_4}T)^2, 
\label{tadpoleeq}
\ee
where $T$ is the contribution of the tadpole diagram. 
The equation \eqref{tadpoleeq} is quadratic so there are two solutions in general. We restrict our interest to the region of coupling constants where 
\be 
\sgn(y_4) =  \sgn(y_4+3 c_F y_6(\lambda -\sgn(\lambda))),
\label{positivity} 
\ee
whose physical implication will be discussed in the next section. 
We choose a solution which has correct perturbative continuation so that $T\to {k_F \over 4\pi}y_2^2$ with $y_6, \lambda \to0$: 
\beal{
T=&\frac{-k_Fy_4^2}{6 \pi y_6} \(1 - \sqrt{1 + {3 y_6 \over y_4^2} \left(c_F^2 (|\lambda| -2) |\lambda| +y_2^2\right) }\) 
={k_F \over 2\pi} c_F y_4 (\lambda -\text{sgn}(\lambda ))
\label{T}
}
where in the second equality we used the gap equation \eqref{cf} to remove $y_2^2$. 

%%%%%%%%%%%%%%%%%%%%%%%%%%%%%%%%%%%%%%%%%
\section{ Four point function of fermions }
\label{4pt} 

\begin{figure}[thbp] 
  \begin{center}
  \subfigure[]{\includegraphics[scale=.6]{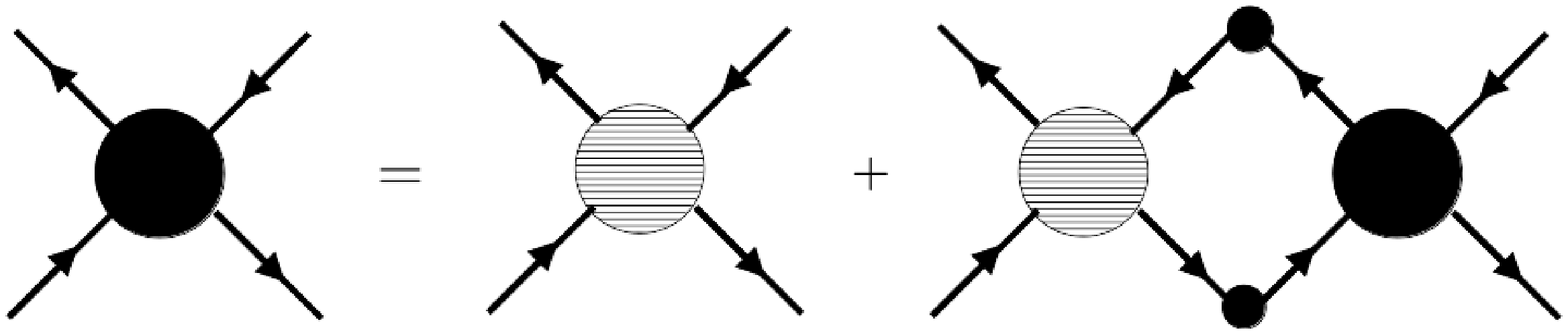}
% \label{selfenergy}
  }
  \qquad\qquad
  \subfigure[]{\includegraphics[scale=.6]{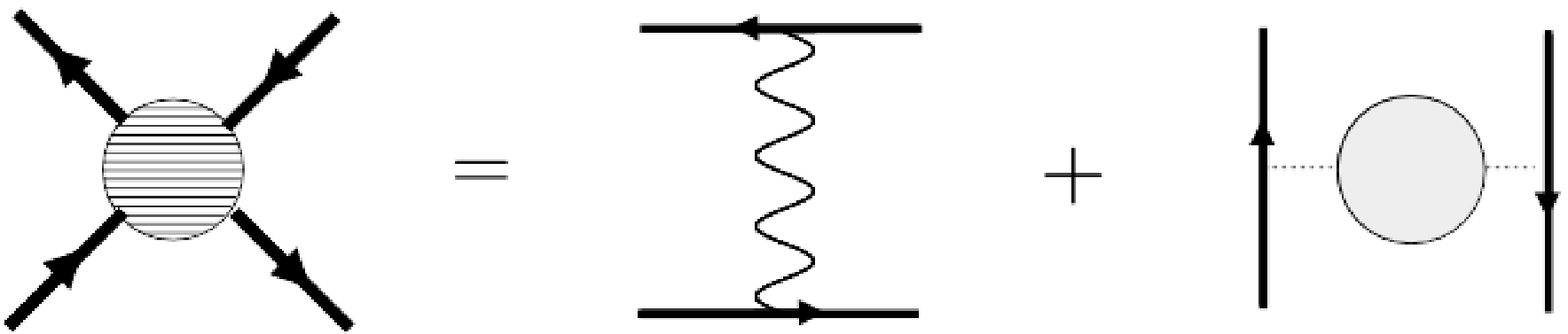}
% \label{tadpole}
  }
  \qquad\qquad
  \subfigure[]{\includegraphics[scale=.6]{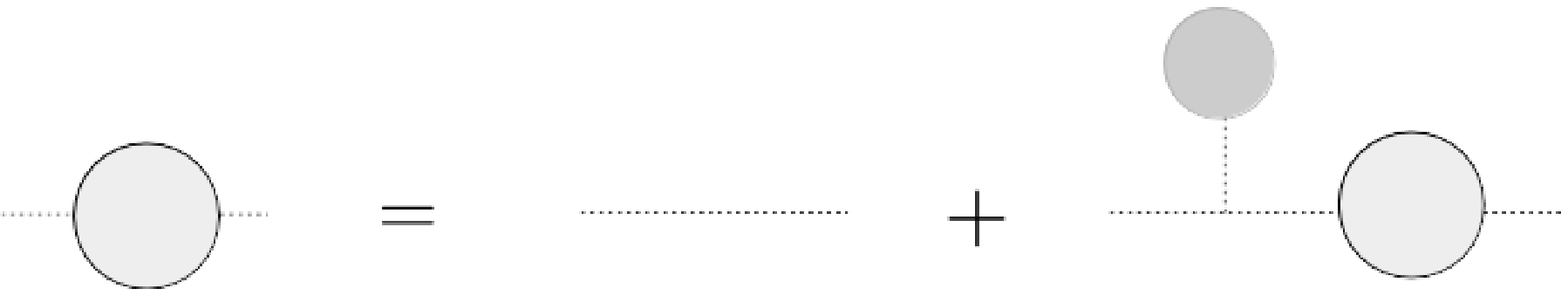}
% \label{tadpole}
  }
  \end{center}
  \vspace{-0.5cm}
  \caption{(a) shows the exact bootstrap diagram for the four point function of fermions in the large $N$ limit. Here the black bubble with two or four lines represents the exact two or four point function, respectively, and the striped bubble represents the ladder bar of this ladder diagram. (b) shows the definition of the ladder bar, in which the striped bubble is given by summation of one {\it gluon} exchange diagram and that with fermionic lines connected by the auxiliary fields with a gray bubble, which is defined as the two point function whose 1PI diagram is given only by the tadpole as depicted in (c).}
\label{4ptfeynmandiagram}
\end{figure}

In this section we compute four point function of fermions in the large $N$ limit. 
Since amputated connected four point function is equivalent to four point vertex, we compute the latter. 
In particular we compute the four point vertex of the large $N$ form such that 
\be 
-\half \int{d^3 p\over(2\pi)^3}{d^3 k\over(2\pi)^3}{d^3 q\over(2\pi)^3} \bar\psi_{\alpha',m}(-p-q) \psi^{\beta',m}(p) F^{\alpha'}\!_{\beta'}\!^{\beta}\!_{\alpha} (p, k; q) \bar\psi_{\beta,n}(-k) \psi^{\alpha,n}(k+q),
\ee
where the spinor indices denoted by Greek letters are explicitly restored.
Under the large $N$ limit the bootstrap diagram of the four point vertex is  simplified to be self-consistent without any other higher point vertices and given by the ladder diagram as described in Figure \ref{4ptfeynmandiagram}. 
From the figure \ref{4ptfeynmandiagram} the Schwinger-Dyson equation for the four point vertex reads 
\beal{ 
&F^{\alpha'}\!_{\beta'}\!^{\beta}\!_{\alpha} (p, k; q) =F^0\!^{\alpha'}\!_{\beta'}\!^{\beta}\!_{\alpha} (p, k; q)  \nn
&+ N \int {d^3 k' \over (2\pi)^3} F^0\!^{\alpha'}\!_{\beta'}\!^{\beta''}\!_{\alpha''}(p, k'; q)  \alpha_F^{\alpha''}\!_{\alpha'''}(q+k') \alpha_F^{\beta'''}\!_{\beta''} (k')  
F^{\alpha'''}\!_{\beta'''}\!^{\beta}\!_{\alpha}  (k', k; q), 
\label{ffff}
}
where $F^0$ describes the contribution of the ladder piece given by 
\beal{
F^0\!^{\alpha'}\!_{\beta'}\!^\beta\!_\alpha (p, k; q) 
=& i\gamma^\mu\!^\beta\!_{\beta'} i\gamma^\nu\!^{\alpha'}\!_\alpha G_{\nu\mu} (-p+k) 
+ {2\pi \over  k_F \wt y_4 } \delta_{\beta'}^{\alpha'} \delta ^\beta_\alpha
\label{F0}
}
with 
\beal{ 
G_{\mu\nu}(p) =& {2\pi \over ik_Fp_-}(\delta_{\mu,+}\delta_{\nu,3}-\delta_{\mu,3}\delta_{\nu,+}), 
\label{gaugepropagator} \\
\wt y_4 =&y_4 + 3c_F y_6(\lambda-\sgn(\lambda)). 
\label{wty4}
}
We give some comments. The first term in \eqref{F0} is the contribution of one gluon exchange diagram depicted in Figure \ref{4ptfeynmandiagram}(b). $G_{\mu\nu}(p)$ is the Chern-Simons gauge propagator, which is given as \eqref{gaugepropagator} in the light-cone gauge, where the matter loop correction is subleading in $1/N$ expansion.
The second term in \eqref{F0} describes the contribution from the auxiliary field. It is given by the exact two point correlator of the auxiliary field excluding fermion loops, which are already taken into account in \eqref{ffff} described in Figure \ref{4ptfeynmandiagram}(a). 
Denoting the contribution by ${- 2\pi \over  k_F \wt y_4 }$, we can determine it by the bootstrap diagram given by Figure \ref{4ptfeynmandiagram}(c), which reads 
\be 
{- 2\pi \over  k_F \wt y_4 } = {- 2\pi \over  k_F y_4 }+ (-3! {k_F \over 4\pi}y_6) \times ({- 2\pi \over  k_F y_4 })\times T \times ({- 2\pi \over  k_F y_4 })\times ({- 2\pi \over  k_F \wt y_4 })
\ee
where $T$ is the contribution of exact tadpole diagram given by \eqref{T}. 
Solving this gives \eqref{wty4}. Therefore in the coupling region specified by \eqref{positivity} the quantum two point correlator of the scalar field (excluding fermion loops) has the same sign as that of tree level, which may suggest that the current system is reflection positive or stable in the quantum mechanical sense under the coupling region. 

To solve the bootstrap equation \eqref{ffff} it is convenient to transform the spinor indices into the vector ones by using the gamma matrices so that 
\be
F^{\alpha'}\!_{\beta'}\!^{\beta}\!_{\alpha} (p, k; q) 
=F_{\bar\mu\bar\nu} (p, k; q)  \gamma^{\bar\mu}\!^{\alpha'}\!_{\beta'}\gamma^{\bar\nu}\!^{\beta}\!_{\alpha}
\ee
where $\bar\mu=1,2,3,I$ and $\gamma^I= I$. 
Plugging this into the above and straight-forward gamma gymnastics we find that the above equation reduces to two sets of consistent integral equations. 
One set is on $F_{II}$ and $F_{+I}$, which is given by
\beal{
&F_{II}(p, k) = {2\pi \over k_F\wt y_4} -2 \pi i \lambda  \int {d^3 k' \over (2\pi)^3} 
{2 \(2 F_{+I} (k',k) k'\!_-^2 + F_{II} (k',k)  k'\!_- (2 i \Sigma_{I}(k')-q_3)\) \over ( -p +k')_- (k'^2 +c_F^2) ((k'+q)^2 +c_F^2)} \nn
& + {2\pi\lambda \over \wt y_4} \int {d^3 k' \over (2\pi)^3} 
{-2 \( F_{+I} (k',k) k'\!_- (2 i \Sigma_{I}(k')+q_3) + F_{II} (k',k) (k_s'^2 + c_F^2 -2 \Sigma_{I}^2(k')+ k_3'^2 + q_3 k_3 '\)  \over  (k'^2 +c_F^2) ((k'+q)^2 +c_F^2)},
\label{FII} \\
&F_{+I}(p, k) = {- \pi i \over 2k_F (-p+k)_-} -2 \pi i \lambda \times  \nn
&\int {d^3 k' \over (2\pi)^3} 
{-2 \( F_{+I} (k',k) k'\!_- (2 i \Sigma_{I}(k')+q_3) + F_{II} (k',k) (k_s'^2 + c_F^2 -2 \Sigma_{I}^2(k')+ k_3'^2 + q_3 k_3 '\) \over ( -p +k')_- (k'^2 +c_F^2) ((k'+q)^2 +c_F^2)}, 
\label{F+I} 
}
where we abbreviate the third variable $q$ in the scattering function. 
The other set is on $F_{I+}$ and $F_{++}$, which is given by
\beal{
&F_{I+}(p, k) = {2 \pi i \over k_F (-p+k)_-} -2 \pi i \lambda  \int {d^3 k' \over (2\pi)^3} 
{2 \(2 F_{++} (k',k) k'\!_-^2 + F_{I+} (k',k)  k'\!_- (2 i \Sigma_{I}(k')-q_3)\) \over ( -p +k')_- (k'^2 +c_F^2) ((k'+q)^2 +c_F^2)} \nn
& + {2\pi\lambda \over \wt y_4} \int {d^3 k' \over (2\pi)^3} 
{-2 \( F_{++} (k',k) k'\!_- (2 i \Sigma_{I}(k')+q_3) + F_{I+} (k',k) (k_s'^2 + c_F^2 -2 \Sigma_{I}^2(k')+ k_3'^2 + q_3 k_3 '\)  \over  (k'^2 +c_F^2) ((k'+q)^2 +c_F^2)}, 
\label{FI+} \\
&F_{++}(p, k) = 4 \pi i \lambda \int {d^3 k' \over (2\pi)^3} 
{ F_{++} (k',k) k'\!_- (2 i \Sigma_{I}(k')+q_3) + F_{II} (k',k) (k_s'^2 + c_F^2 -2 \Sigma_{I}^2(k')+ k_3'^2 + q_3 k_3 ') \over ( -p +k')_- (k'^2 +c_F^2) ((k'+q)^2 +c_F^2)}. 
\label{F++} 
}
These integral equations can be solved exactly in a specific momentum frame such that 
\be 
q_\pm := {q_1 \pm i q_2 \over \sqrt2} = 0.  
\label{com}
\ee
This frame matches the center of mass frame in S-channel scattering, which is investigated in the next section. 
Then let us set the following ansatz: 
\beal{
F_{II}(p, k) =& { p_- \over (-p+k)_- } B_1 (p_s, k_s) +B_2 (p_s, k_s), 
\label{ansatzFII}\\
F_{+I}(p, k) =& { 1 \over (-p+k)_- } A_1 (p_s, k_s) + { 1 \over p_- } A_2 (p_s, k_s),
\label{ansatzF+I}
}
and 
\beal{
F_{I+}(p, k) =& { 1 \over (-p+k)_- } B_3 (p_s, k_s) + { 1 \over k_- } B_4 (p_s, k_s), 
\label{ansatzFI+}\\
F_{++}(p, k) =& { 1 \over p_- (-p+k)_- } A_3 (p_s, k_s) +{ 1 \over p_-  k_- } A_4 (p_s, k_s). 
\label{ansatzF++}
}
Plugging this ansatz into the above and performing integral for $k'_3$ and angular part of $k'_-$ we obtain
\beal{
B_1 (p_s, k_s)=&- 4 \pi i \lambda \int_{h_p}^{h_k} {d h \over (2\pi)} 
{ 2 A_1 + B_1 (2 i \Sigma_{I} -q_3 ) \over 4 h^2 +q_3^2},
\label{B1}\\
A_1 (p_s, k_s)=&{ -2 \pi i \over k_F} + 4 \pi i \lambda \int_{h_p}^{h_k} {d h \over (2\pi)} 
{ A_1 (2 i \Sigma_{I} + q_3 ) +2 B_1 (h^2 - \Sigma_{I}^2 ) \over 4 h^2 +q_3^2},
\label{A1}\\
B_2 (p_s, k_s)=& {2\pi \over k_F\wt y_4} - 4 \pi i \lambda \(- \int_{h_k}^{\infty} {d h \over (2\pi)} 
{ 2 A_1 + B_1 (2 i \Sigma_{I} -q_3 ) \over 4 h^2 +q_3^2} 
+\int_{h_p}^{\infty} {d h \over (2\pi)} 
{ 2 A_2 + B_2 (2 i \Sigma_{I} -q_3 ) \over 4 h^2 +q_3^2}\) \nn
& - {4\pi\lambda \over \wt y_4} \bigg(- \int_{h_k}^{\infty} {d h \over (2\pi)} 
{A_1 (2 i \Sigma_{I} + q_3 ) +2 B_1 (h^2 - \Sigma_{I}^2 ) \over 4 h^2 +q_3^2} \nn
&+\int_{h_0}^{\infty} {d h \over (2\pi)} 
 {A_2 (2 i \Sigma_{I} + q_3 ) +2 B_2 (h^2 - \Sigma_{I}^2 ) \over 4 h^2 +q_3^2}\bigg), 
\label{B2} \\
A_2 (p_s, k_s)=&- 4 \pi i \lambda  \int_{h_0}^{h_p} {d h \over (2\pi)} 
{ A_2 (2 i \Sigma_{I} + q_3 ) +2 B_2 (h^2 - \Sigma_{I}^2 ) \over 4 h^2 +q_3^2}, 
\label{A2}
}
and
\beal{
B_3 (p_s, k_s)=&{ 2 \pi i \over k_F}- 4 \pi i \lambda \int_{h_p}^{h_k} {d h \over (2\pi)} 
{ 2 A_3 + B_3 (2 i \Sigma_{I} -q_3 ) \over 4 h^2 +q_3^2},
\label{B3}\\
A_3 (p_s, k_s)=& 4 \pi i \lambda \int_{h_p}^{h_k} {d h \over (2\pi)} 
{ A_3 (2 i \Sigma_{I} + q_3 ) +2 B_3 (h^2 - \Sigma_{I}^2 ) \over 4 h^2 +q_3^2},
\label{A3}\\
B_4 (p_s, k_s)=&- 4 \pi i \lambda \int_{h_p}^{\infty} {d h \over (2\pi)} 
{ 2 A_4 + B_4 (2 i \Sigma_{I} -q_3 ) \over 4 h^2 +q_3^2} - {4\pi\lambda \over \wt y_4} \bigg(\int^{h_k}_{h_0} {d h \over (2\pi)} 
{A_3 (2 i \Sigma_{I} + q_3 ) +2 B_3 (h^2 - \Sigma_{I}^2 ) \over 4 h^2 +q_3^2} \nn
&+\int_{h_0}^{\infty} {d h \over (2\pi)} 
 {A_4 (2 i \Sigma_{I} + q_3 ) +2 B_4 (h^2 - \Sigma_{I}^2 ) \over 4 h^2 +q_3^2}\bigg), 
\label{B4}\\
A_4 (p_s, k_s)=&- 4 \pi i \lambda \biggl(\int^{h_k}_{h_0} {d h \over (2\pi)} 
{A_3 (2 i \Sigma_{I} + q_3 ) +2 B_3 (h^2 - \Sigma_{I}^2 )  \over 4 h^2 +q_3^2} \nn
&+\int^{h_p}_{h_0} {d h \over (2\pi)} 
{ A_4 (2 i \Sigma_{I} + q_3 ) +2 B_4 (h^2 - \Sigma_{I}^2 )  \over 4 h^2 +q_3^2}\biggl),
\label{A4}
}
where we set $h_p:= \sqrt{p_s^2+c_F^2}$, and in the right-hand side $\Sigma_I=\lambda h +(\sgn(\lambda)-\lambda)c_F$ and $A_i=A_i(k_s',k_s), B_i=B_i(k_s',k_s)$ with $h=h_{k'}$, $i=1,2,3,4$.   

In order to solve this let us take derivative with respect to $h_p$ for both sides. Then we obtain 
\beal{
{\partial B_i \over \partial h_p} =& 2 i \lambda { 2 A_i + B_i (2 i \Sigma_{I}(p) -q_3 ) \over 4 h_p^2 +q_3^2}
\label{Bip}\\
{\partial A_i \over \partial h_p} =&- 2 i \lambda { A_i (2 i \Sigma_{I}(p) + q_3 ) +2 B_i (h_p^2 - \Sigma_{I}(p)^2 ) \over 4 h_p^2 +q_3^2}
\label{Aip}
}
for all $i=1,2,3,4$. 
From these two equations one can derive
\be
{\partial  \over \partial h_p}  ( B_i (2i \Sigma +q_3) + 2 A_i ) =0. 
\ee
Thus $A_i$ can be related to $B_i$ by
\be
A_i(p_s,k_s)  = - B_i(p_s,k_s) (i \Sigma(p) +\half q_3) + \alpha_i
\label{AtoB} 
\ee
where $\alpha_i$ is an unknown function independent of $p_s$. 
Plugging this back into \eqref{Bip} we find 
\be
{\partial B_i \over \partial h_p} = 2 i \lambda { -2 q_3 B_i + 2 \alpha_i \over 4 h_p^2 +q_3^2}. 
\ee
This can be solved as 
\be
B_i (p_s, k_s) = {\alpha_i \over q_3 } + \beta_i f(p_s) 
\label{Bsol}
\ee
where $\beta_i$ is another integration {\it constant} independent of $p_s$ and $ f(p_s) $ is defined by 
\be
f(p_s):= \exp\[ -2i \lambda \arctan { 2 \sqrt {p_s^2 + c_F^2} \over q_3 } \].
\label{f} 
\ee 
The integration constants $\alpha_i, \beta_i$ can be determined so as to satisfy the following conditions derived from the integral equations.  
\beal{ 
\lim_{p_s\to k_s} B_1(p_s,k_s) =& 0, 
  \label{B1condition} \\
\lim_{p_s\to k_s} A_1(p_s,k_s) =& {- 2 \pi i \over k_F}, 
  \label{A1condition} \\
\lim_{p_s\to \infty} B_2(p_s,k_s) =& {2\pi \over k_F\wt y_4}+ 4 \pi i \lambda \int_{h_k}^{\infty} {d h \over (2\pi)} 
{ 2 A_1 + B_1 (2 i \Sigma_{I} -q_3 ) \over 4 h^2 +q_3^2} \nn
& - {4\pi\lambda \over \wt y_4} \bigg(- \int_{h_k}^{\infty} {d h \over (2\pi)} 
{A_1 (2 i \Sigma_{I} + q_3 ) +2 B_1 (h^2 - \Sigma_{I}^2 ) \over 4 h^2 +q_3^2} \nn
&+\int_{h_0}^{\infty} {d h \over (2\pi)} 
 {A_2 (2 i \Sigma_{I} + q_3 ) +2 B_2 (h^2 - \Sigma_{I}^2 ) \over 4 h^2 +q_3^2}\bigg), 
 \label{B2condition} \\
\lim_{p_s\to 0} A_2(p_s,k_s) =& 0, 
  \label{A2condition} 
}
and 
\beal{ 
\lim_{p_s\to k_s} B_3(p_s,k_s) =& {2 \pi i \over k_F},
  \label{B3condition} \\
\lim_{p_s\to k_s} A_3(p_s,k_s) =& 0, 
  \label{A3condition} \\
\lim_{p_s\to \infty} B_4(p_s,k_s) =&- {4\pi\lambda \over \wt y_4} \bigg(\int_{h_0}^{h_k} {d h \over (2\pi)} 
{A_3 (2 i \Sigma_{I} + q_3 ) +2 B_3 (h^2 - \Sigma_{I}^2 ) \over 4 h^2 +q_3^2} \nn
&+\int_{h_0}^{\infty} {d h \over (2\pi)} 
 {A_4 (2 i \Sigma_{I} + q_3 ) +2 B_4 (h^2 - \Sigma_{I}^2 ) \over 4 h^2 +q_3^2}\bigg), 
  \label{B4condition} \\
\lim_{p_s\to 0} A_4(p_s,k_s) =& - 4 \pi i \lambda \int^{h_k}_{h_0} {d h \over (2\pi)} 
{ A_3 (2 i \Sigma_{I} + q_3 ) +2 B_3 (h^2 - \Sigma_{I}^2) \over 4 h^2 +q_3^2}. 
  \label{A4condition} 
}
By using \eqref{A1condition}, \eqref{B1condition}, \eqref{A3condition}, \eqref{B3condition}, $\alpha_1,\beta_1, \alpha_3, \beta_3$ can be easily determined as   
\beal{ 
\alpha_1=&  {- 2 \pi i \over k_F}, \quad \beta_1 = { 2 \pi i \over k_F q_3 f(k_s)}, \\
\alpha_3=&   (2 i \Sigma_{I}(k_s) +q_3) {\pi i \over k_F}, \quad 
\beta_3 = { \pi i (-2 i \Sigma_{I}(k_s)+q_3) \over k_F q_3 f(k_s)}.
}
On the other hand, the right-hand sides in \eqref{B2condition}, \eqref{B4condition} contain divergent integrals, which need regularization.  This divergence originates in non-renormalizable interaction  
in the sense of power counting in \eqref{csfnonlinear2}.
We regularize the divergence by dimensional regularization demonstrated in detail in \cite{Jain:2012qi}, which is skipped here and move it to Appendix \ref{regularization}.  
After the regularization we can determine $\alpha_2,\beta_2, \alpha_4, \beta_4$ by using \eqref{A2condition}, \eqref{B2condition}, \eqref{A4condition}, \eqref{B4condition} as 
\beal{
\alpha_i = {(num)_{\alpha_i} \over (den)_{\alpha_i}}, \quad 
\beta_i = {(num)_{\beta_i} \over (den)_{\beta_i}}, \quad 
}
where 
\beal{
\hspace{-2cm}
(num)_{\alpha_2}=& 2 \pi  f(0) (q_3+2 i \Sigma_I(0)) (f(k_s) (2 c_F (\lambda-\sgn(\lambda))-i q_3-2 \wt y_4) \nn
& +f(\infty ) (-2 c_F (\lambda-\sgn(\lambda))-i q_3+2 \wt y_4)),\\
(den)_{\alpha_2}=& k_F f(k_s) \{f(\infty ) (2 \Sigma_I(0)+i q_3) (2 c_F (\lambda-\sgn(\lambda))+i q_3-2 \wt y_4) \nn
&+f(0) \left(c_F (-4 (\lambda-\sgn(\lambda)) \Sigma_I(0)+2 i (\lambda+\sgn(\lambda)) q_3)+q_3^2-2 i q_3 (\Sigma_I(0)+\wt y_4)+4 \Sigma_I(0) \wt y_4\right)\},\\
\hspace{-2cm}
(num)_{\beta_2}=&-2 \pi  (q_3-2 i \Sigma_I(0)) (f(k_s) (-2 c_F (\lambda-\sgn(\lambda))+i q_3+2 \wt y_4)\nn
&+f(\infty ) (2 c_F (\lambda-\sgn(\lambda))+i q_3-2 \wt y_4)),\\
(den)_{\beta_2}=&k_F q_3 f(k_s) \{f(\infty ) (2 \Sigma_I(0)+i q_3) (2 c_F (\lambda-\sgn(\lambda))+i q_3-2 \wt y_4) \nn
&+ f(0) \left(c_F (-4 (\lambda-\sgn(\lambda)) \Sigma_I(0)+2 i (\lambda+\sgn(\lambda)) q_3)+q_3^2-2 i q_3 (\Sigma_I(0)+\wt y_4)+4 \Sigma_I(0) \wt y_4\right)\}, \\
(num)_{\alpha_4}=&\pi  (f(k_s) (q_3-2 i \Sigma_I(0)) (q_3+2 i \Sigma_I(k_s))-f(0) (q_3+2 i \Sigma_I(0)) (q_3-2 i \Sigma_I(k_s))) \times \nn
& (2 (c_F-\Sigma_I(0)) f(0)+f(\infty ) (2 c_F (\lambda-\sgn(\lambda))+i q_3-2 \wt y_4)),\\
(den)_{\alpha_4}=&k_F f(k_s) \{ f(\infty ) (2 \Sigma_I(0)+i q_3) (2 c_F (\lambda-\sgn(\lambda))+i q_3-2 \wt y_4) \nn
& +f(0) \left(c_F (-4 (\lambda-\sgn(\lambda)) \Sigma_I(0)+2 i (\lambda+\sgn(\lambda)) q_3)+q_3^2-2 i q_3 (\Sigma_I(0)+\wt y_4)+4 \Sigma_I(0) \wt y_4\right)\},\\
\hspace{-2cm}
(num)_{\beta_4}=&\pi  (-2 (-c_F \lambda +\Sigma_I(0)+\wt y_4)-i q_3) (f(0) (q_3+2 i \Sigma_I(0)) (q_3-2 i \Sigma_I(k_s)) \nn
&-f(k_s) (q_3-2 i \Sigma_I(0)) (q_3+2 i \Sigma_I(k_s))), \\
(den)_{\beta_4}=&k_F q_3 f(k_s) \{f(\infty ) (2 \Sigma_I(0)+i q_3) (2 c_F (\lambda-\sgn(\lambda))+i q_3-2 \wt y_4) \nn
& +f(0) \left(c_F (-4 (\lambda-\sgn(\lambda)) \Sigma_I(0)+2 i (\lambda+\sgn(\lambda)) q_3)+q_3^2-2 i q_3 (\Sigma_I(0)+\wt y_4)+4 \Sigma_I(0) \wt y_4\right)\}.
}

Comments are in order. 
Firstly by taking the limit with $\wt y_4\to\infty$ this solution reduces to that obtained in \cite{Jain:2014nza} up to the convention adopted there, where the fermionic four point function in massive free fermion theory gauged by Chern-Simons interaction was computed. This is consistent with the fact that the Schwinger-Dyson equation \eqref{ffff} with \eqref{F0} boils down to that of the Chern-Simons fermion vector model.

Secondly by sending $\lambda, {y_2^2\over\lambda}, {y_6\over\lambda} \to0$ with ${2\pi \lambda \over y_4} =g^F_4$ fixed this solution reduces to the four point function of Gross-Neveu model computed in \cite{Rosenstein:1988pt}.     
Indeed under this limit all the components of $F_{\bar\mu\bar\nu}$ vanish except $F_{II}$, which is given by 
\beal{ 
F_{II}=&{1 \over N}  { {g^F_4 }  \over 1 +  g^F_4 \( -{ c_F \over 2 \pi}  -( q_3^2 + 4c_F^2) {- \tan^{-1} {2c_F \over q_3} +  \tan^{-1} {2 \infty \over q_3} \over 4\pi q_3} \)} \nn
=&{4\pi \over N}  {q_3  \over   -( q_3^2 + 4c_F^2)  \tan^{-1} ({q_3\over  2c_F})} 
% =&{1 \over N}  {4\pi q^0  \over ( (q^0)^2 - 4c_F^2) ( \tanh^{-1} {2c_F \over q^0} + {i\pi \over 2} ) }
\label{4ptGN}
}
where in the second equation we used $c_F = {2\pi \over g_4^F}$ under the limit and the formula $\tan^{-1}(x)+\tan^{-1}({1\over x}) = \tan^{-1}(\infty)$. Covariantizing \eqref{4ptGN} by exchanging $q_3 \to \sqrt{(q_\mu)^2}$ gives (9) in \cite{Rosenstein:1988pt} up to an overall numerical factor.

\section{S-matrix and duality}
\label{Smatrix} 

In this section we compute two body scattering matrix of fermions in the general Chern-Simons fermion vector model in the 't Hooft large $N$ limit. For this we apply the method for computing S-matrix developed in \cite{Jain:2014nza}, in which the fermionic S-matrix in regular Chern-Simons fermion vector model was computed. 

In the current system there are two kinds of scattering process to be studied, that is particle-particle scattering and particle-antiparticle one.%
\footnote{Antiparticle-antiparticle S-matrix can be obtained by charge conjugation of particle-particle one. 
}
We focus on the particle-antiparticle scattering in this paper since the particle-particle S-matrix is to be obtained by analytic continuation from particle-antiparticle S-matrix as shown in \cite{Jain:2014nza}. 
Asymptotic states for this process are constructed in the appendix \ref{states}. 
We consider particle-antiparticle scattering process such that 
\be
\ket{\vec p_3, -, k; \vec p_2, +, j} \to \ket{ -\vec p_4, -,l; -\vec p_1, +, i}. \quad 
\ee
This means that the $k$-the particle with momentum $\vec p_3$ and the $j$-th antiparticle with momentum $\vec p_2$ come from the past infinity and go into the $l$-the particle with momentum $-\vec p_4$ and the $i$-th antiparticle with momentum $-\vec p_1$ at future infinity. 
By taking into account the momentum conservation the S-matrix of this process can be written as 
\bes{
&\bra{ -\vec p_4, -,l; -\vec p_1, +, i}\hat S\ket{\vec p_3, -, k; \vec p_2, +, j} \\ 
=& S(-\vec p_4,-,l;-\vec p_1,+,i|\vec p_3,-,k;\vec p_2,+,j) (2\pi)^3\delta^3(p_1+p_2+p_3+p_4). 
\label{Smatrix}
}
Furthermore the S-matrix can be decomposed in terms of the gauge indices as \cite{Jain:2014nza} 
\beal{
&S(-\vec p_4,-,l;-\vec p_1,+,i|\vec p_3,-,k;\vec p_2,+,j) \nn
=&(\delta_k^l \delta_i^j - {1\over N} \delta_i^l \delta_k^j) S_T(-\vec p_4,-;-\vec p_1,+|\vec p_3,-;\vec p_2,+)
+{1\over N}{\delta_k^j \delta_i^l } S_S(-\vec p_4,-;-\vec p_1,+|\vec p_3,-;\vec p_2,+). 
\label{SmatrixST}
}
We call the coefficients $S_T(-\vec p_4,-;-\vec p_1,+|\vec p_3,-;\vec p_2,+), S_S(-\vec p_4,-;-\vec p_1,+|\vec p_3,-;\vec p_2,+)$ T-channel (adjoint channel) S-matrix, S-channel (singlet channel) one, respectively. 
Similarly the transition matrix defined by 
$\hat S =\hat 1 + i \hat T$ is decomposed as  
\beal{ 
&T(-\vec p_4,-,l;-\vec p_1,+,i|\vec p_3,-,k;\vec p_2,+,j) \nn
&=(\delta_k^l \delta_i^j - {1\over N} \delta_i^l \delta_k^j)T_T(-\vec p_4,-;-\vec p_1,+|\vec p_3,-;\vec p_2,+)
+{1\over N}{\delta_k^j \delta_i^l }T_S(-\vec p_4,-;-\vec p_1,+|\vec p_3,-;\vec p_2,+).
\label{TmatrixST}
}
Then S-matrix for each channel satisfies 
\beal{
&S_T(-\vec p_4,-;-\vec p_1,+|\vec p_3,-;\vec p_2,+) \nn
= & {\bra{ -\vec p_4, -; -\vec p_1, +} \vec p_3, -; \vec p_2, +\rangle \over (2\pi)^3\delta^3(p_1+p_2+p_3+p_4)} +iT_T(-\vec p_4,-;-\vec p_1,+|\vec p_3,-;\vec p_2,+), 
\label{SmatrixT}\\
&S_S(-\vec p_4,-;-\vec p_1,+|\vec p_3,-;\vec p_2,+) \nn
=&{\bra{ -\vec p_4, -; -\vec p_1, +} \vec p_3, -; \vec p_2, +\rangle \over (2\pi)^3\delta^3(p_1+p_2+p_3+p_4)} +iT_S(-\vec p_4,-;-\vec p_1,+|\vec p_3,-;\vec p_2,+),
\label{SmatrixS}
}
where
\be
\bra{ -\vec p_4, -; -\vec p_1, +} \vec p_3, -; \vec p_2, +\rangle = (2\pi)^2 2 E_{\vec p_3} \delta^2(\vec p_3+\vec p_4) (2\pi)^2 2 E_{\vec p_2} \delta^2(\vec p_2+\vec p_1).
\ee
Under the definition the T-matrix in T-channel is of order $1/N$, while that in S-channel is of order one. 

% \subsection{Unitarity condition} 
% \label{unitarity} 

Let us determine the unitarity condition for each channel. 
Let us start the unitarity condition with respect to the S-matrix 
\be
\hat S^\dagger \hat S=1. 
\label{unitary} 
\ee
In terms of the transition matrix, this is written as  
\be
-i (\hat T -\hat T^\dagger) = \hat T^\dagger \hat T. 
\label{unitary2}
\ee
Let us sandwich both sides inside the two particle states so that 
\be 
{\bra{-\vec p_4,-,l;-\vec p_1,+,i} -i (\hat T- \hat T^\dagger) \ket{\vec p_3,-,k;\vec p_2,+,j} \over (2\pi)^3\delta^3(p_1+p_2+p_3+p_4)}
={ \bra{-\vec p_4,-,l;-\vec p_1,+,i} \hat T^\dagger \hat T \ket{\vec p_3,-,k;\vec p_2,+,j} \over (2\pi)^3\delta^3(p_1+p_2+p_3+p_4)}.
\ee
By using \eqref{TmatrixST} the left-hand side is computed as 
\beal{
{\rm LHS}
=&-i\{ (\delta_k^l \delta_i^j - {1\over N} \delta_i^l \delta_k^j)( T_T(-\vec p_4,-;-\vec p_1,+|\vec p_3,-;\vec p_2,+)
- T_T(\vec p_3,-;\vec p_2,+|-\vec p_4,-;-\vec p_1,+)^* ) \nn
&+{1\over N} \delta_k^j \delta_i^l ( T_S(-\vec p_4,-;-\vec p_1,+|\vec p_3,-;\vec p_2,+)  - T_S(\vec p_3,-;\vec p_2,+|-\vec p_4,-;-\vec p_1,+)^*) \}. 
}
In order to compute the right-hand side we insert identity such that 
\beal{
1 =\sum_{k,j}\int {d^3 r_1 \over (2\pi)^3}{d^3 r_2 \over (2\pi)^3} 
(2\pi) \delta(r_1^2+ c_F^2) \theta(r_1^0) (2\pi) \delta(r_2^2+ c_F^2)\theta(r_2^0) \nn \ket{\vec r_1,-,k;\vec r_2,+,j} \bra{\vec r_1,-,k;\vec r_2,+,j} + \cdots
}
where the ellipsis contains other many-particle states more than 2-particle one, which is suppressed by $1/N$ expansion. 
Then the right-hand side is computed as 
\beal{
{\rm RHS}
=&\int {d^3 r_1 \over (2\pi)^3}{d^3 r_2 \over (2\pi)^3} (2\pi) \delta(r_1^2+ c_{B}^2) \theta(r_1^0) (2\pi) \delta(r_2^2+ c_{B}^2)\theta(r_2^0) \nn
&\{ (\delta_{k}^l \delta_i^{j}-{\delta_{i}^l \delta_k^{j}\over N}) T_T(-\vec p_4,-;-\vec p_1,+|\vec r_1,-;\vec r_2,+)T_T(\vec p_3,-;\vec p_2,+|\vec r_1,-;\vec r_2,+)^* \nn
&+{ \delta_{k}^{j} \delta^l_i \over N} T_S(-\vec p_4,-;-\vec p_1,+|\vec r_1,-;\vec r_2,+) T_S(\vec p_3,-;\vec p_2,+|\vec r_1,-;\vec r_2,+)  ^* \}.
}
Recalling the fact that the transition matrix in T-channel is of order $1/N$, the product of $T_T$ and $T_T^*$ is of order $1/N^2$, which is subleading in $1/N$ expansion. 
Thus at the leading order of large $N$ limit the unitarity condition is given by 
\beal{
&-i\{ T_T(-\vec p_4,-;-\vec p_1,+|\vec p_3,-;\vec p_2,+)
- T_T(\vec p_3,-;\vec p_2,+|-\vec p_4,-;-\vec p_1,+)^*\} = 0 
\label{unitarityTchannel}
}
for T-channel, and 
\beal{
&-i\{ T_S(-\vec p_4,-;-\vec p_1,+|\vec p_3,-;\vec p_2,+)
- T_S(\vec p_3,-;\vec p_2,+|-\vec p_4,-;-\vec p_1,+)^*\}  \nn
=&\int {d^3 r_1 \over (2\pi)^3}{d^3 r_2 \over (2\pi)^3} (2\pi) \delta(r_1^2+ c_F^2) \theta(r_1^0) (2\pi) \delta(r_2^2+ c_F^2)\theta(r_2^0) \nn
& T_S(-\vec p_4,-;-\vec p_1,+|\vec r_1,-;\vec r_2,+) T_S(\vec p_3,-;\vec p_2,+|\vec r_1,-;\vec r_2,+)  ^* 
\label{unitaritySchannel}
}
for S-channel. 

\subsection{T-channel} 
\label{Tchannel} 

We determine the transition matrix in T-channel, which can be done by applying LSZ formula to the four point vertex of fermion determined in the previous section. 
Since the exact four point vertex was determined in Euclidean space, we perform Wick rotation such that
\be
x^2 = i x^0, \quad q^2 = i q^0.
\label{wickTchannel}
\ee
In the momentum assignment in the previous section T-channel is given by   
\beal{
-k^0>0, \, p^0>0, \, (k+q)^0<0, \, -(p+q)^0<0.
}
Then the LSZ formula tells us contraction rule of the wave functions associated with the external legs to obtain T-channel transition matrix as follows.  
\beal{
T_T=&T_T( -\vec k-\vec q, -; \vec p+\vec q, +| -\vec k, -; \vec p, +) \nn
=&v^{\alpha}(-\vec k-\vec q) \bar v_{\beta} (-\vec k) F^{\alpha'}{}_{\beta'}{}^{\beta}{}_{\alpha} (\vec p, \vec k, \vec p+\vec q) u^{\beta'}(\vec p) \bar u_{\alpha'} (\vec p+\vec q)\nn
=&[\bar u(\vec p+\vec q) \gamma^{\bar\nu} u(\vec p)] [\bar v (-\vec k)\gamma^{\bar\mu} v(-\vec k-\vec q)] F_{\bar\nu\bar\mu} (\vec p, \vec k, \vec p+\vec q) 
}
where $u(\vec p), v(\vec p)$ are given in \eqref{u}, \eqref{v}. 
By plugging $F_{\bar\nu\bar\mu}$ determined in the previous section into this we obtain 
\beal{
T_T=&-\frac{4 \pi  q_3}{ik_F} \bigg( z - \frac{f(0)+{(2 \sgn(\lambda)c_{F}+iq_3)  (4 c_{F}   (\lambda-\sgn(\lambda))-4\wt y_4+2   iq_3) \over (2 \sgn(\lambda)c_{F}-iq_3) (4 c_{F}   (\lambda-\sgn(\lambda))-4\wt y_4-2   iq_3)}f(\infty ) }{ f(0) - {(2\sgn(\lambda) c_{F}+iq_3)  (4 c_{F}   (\lambda-\sgn(\lambda))-4\wt y_4+2   iq_3) \over (2 \sgn(\lambda)c_{F}-iq_3) (4 c_{F}   (\lambda-\sgn(\lambda))-4\wt y_4-2   iq_3)}f(\infty )} \bigg) \nn
=& -\frac{4 \pi  q_3}{ik_F} ( z  + i \tan X_F(q_3) )
\label{TchannelTmatrix}
}
where we set 
\beal{
z =& {(p+k)_- \over (p-k)_-}
\label{z} 
}
with $p_-={ p^1+p^0 \over \sqrt2}$, and 
\beal{
X_F(q_3)=&(\lambda-\sgn(\lambda)) \tan ^{-1}\left(\frac{q_3}{2 c_F}\right)-\tan ^{-1}\left(\frac{2 (\wt y_4 - c_F(\lambda-\sgn(\lambda)))}{q_3}\right). 
}

Make some comments. 
Firstly this satisfies the unitary equation \eqref{unitarityTchannel}. This is because the unitarity condition can be written as $T_T = T_T^*|_{q_3\to-q_3}$, which can be easily seen from \eqref{TchannelTmatrix}. 

Secondly by performing duality transformation found in \cite{Minwalla:2015sca} this transition matrix maps that describing two body scattering of scalar quanta in general Chern-Simons scalar theory. 
To see this we recall the duality transformation in \cite{Minwalla:2015sca}. 
\be
k_F=-k_B, \quad \lambda = \lambda_B - \sgn(\lambda_B), \quad
y_6 ={1-x_6 \over 4}, \quad  
y_4 = b_4, \quad y_2^2 = m_B^2,
\label{dualitytransformation}
\ee
which suggests that $c_{F} = c_{B}$. 
Plugging this duality relation into \eqref{TchannelTmatrix} we obtain 
\beal{
T_T
=& {4 \pi q_3 \over i k_B}  \( z + i \tan X_B(q_3)   \) 
}
where 
\beal{
X_B(q_3)=\lambda_B \tan ^{-1}\left(\frac{q_3}{2 c_B}\right)+ \tan ^{-1}\left(\frac{-4 b_4+c_B \lambda_B(1+3 x_6)}{2 q_3}\right). 
\label{XB}
}
This precisely agrees with the transition matrix in T-channel (adjoint channel) in general Chern-Simons bosonic vector model, $T^{(B)}_T$, which is computed in Appendix \ref{bosonicSmatrix} by extending the result in \cite{Jain:2014nza} including the triple trace coupling. 
Thus the S-matrix of this system given by \eqref{SmatrixT} maps that of the dual bosonic system. 

Thirdly one can covariantize this scattering amplitude by rewriting in terms of Mandelstam variables. 
Our definition of Mandelstam variables is%
\footnote{ 
The definition of Mandelstam variables in this paper is different from that  in \cite{Jain:2014nza}. The $s,t,u$ variables there are defined in each channel so that $s$ always becomes positive as adopted in a standard textbook \cite{Peskin:1995ev}, while here they are defined {\it globally} so that $s,t,u$ become positive in S,T,U-channel, respectively. 
}
\beal{
s=&- q^2, \quad 
t=- (p-k)^2, \quad 
u= -(p+k+q)^2, 
\label{stu}
}
which satisfy $s+t+u=4c_F^2$. 
In the frame \eqref{com}, $s = -q_3^2 < 0$. 
Let us consider to rewrite the first term in \eqref{TchannelTmatrix}, which comes from the one gluon exchange diagram. This part can be covariantized as 
\be
-\frac{4 \pi  q_3}{ik_F} z =-\frac{4 \pi }{ik_F} {\varepsilon^{\mu\nu\rho}(p+k)_\mu q_\nu (p-k)_\rho \over (p-k)^2}  
\ee
where $\varepsilon^{013}=1$. 
From a straightforward algebraic calculation we can show that 
\be 
(\varepsilon^{\mu\nu\rho}(p+k)_\mu q_\nu (p-k)_\rho)^2 = stu. 
\ee
which leads to 
\be 
\varepsilon^{\mu\nu\rho}(p+k)_\mu q_\nu (p-k)_\rho = \sigma\sqrt{stu}. 
\ee
where we set $\sigma = \sgn(\varepsilon^{\mu\nu\rho}k_\mu q_\nu p_\rho)$. 
Therefore we find 
\be
-\frac{4 \pi  q_3}{ik_F} z =-\frac{4 \pi }{ik_F} { \sigma\sqrt{stu} \over -t}   =\frac{4 \pi }{k_F} \sigma\sqrt{su \over -t}  
\ee
where we used the fact that $t>0$ in T-channel. 
Substituting these into \eqref{TchannelTmatrix} we obtain the covariant form of the T-channel scattering amplitude as 
\be
T_T= \frac{4 \pi}{k_F} \(\sigma\sqrt{su \over -t} - \sqrt{-s}\tan X_F(\sqrt{-s}) \).
\label{TchannelTmatrixSTU}
\ee

\subsection{S-channel} 
\label{Schannel} 

Finally we determine the transition matrix in S-channel. 
For this let us first compute the transition matrix in S-channel in the same fashion as done in T-channel. 
For S-channel we perform Wick rotation from Euclidean space such that
\be
x^3 = i x^0, \quad q^3 = i q^0.
\label{wickSchannel}
\ee
Then the frame \eqref{com} coincides with the center of mass frame in the two body scattering. 
In the momentum assignment in the previous section S-channel is given by   
\beal{
-(p+q)^0>0, \, p^0>0, \, (k+q)^0<0, \, -k^0<0.
}
In particular $q^0 < 0$. 
From the LSZ formula S-channel transition matrix is obtained by contracting the four point vertex with the wave functions associated with the external legs in such a way that 
\beal{
T_S=&T_S(-\vec k-\vec q, -; \vec k, + | -\vec p-\vec q, -; \vec p, + ) \nn
=&-v^{\alpha}(-k-q) \bar u_{\beta} (k_s) F^{\alpha'}{}_{\beta'}{}^{\beta}{}_{\alpha} (p, k, p+q) u^{\beta'}(p) \bar v_{\alpha'} (-p-q)\nn
=&-[\bar v(-p-q) \gamma^{\bar\nu} u(p)] [\bar u (k_s)\gamma^{\bar\mu} v(-k-q)] F_{\bar\nu\bar\mu} (p, k, p+q) 
}
where $u(\vec p), v(\vec p)$ are given in \eqref{u}, \eqref{v} with double Wick rotation so that firstly $q^0 \to - iq^2$, subsequently $q^3 \to iq^0$.
Note that the minus sign appears when the fermionic fields associated the four point vertex contract with the external states. 
Plugging $F_{\bar\nu\bar\mu}$ determined in the previous section into this gives
\beal{
T_S =& \frac{4 \pi  q^0}{k_F} ( z + \tanh X'_F(-q^0) )
\label{SchannelTmatrix}
}
where $z$ is given by \eqref{z} with $p_-={ p^1-ip^2 \over \sqrt2}$ and 
\beal{
X_F'(-q^0)=&(\lambda-\sgn(\lambda)) \tanh ^{-1}\left(\frac{-q^0}{2 c_F}\right)+\tanh ^{-1}\left(\frac{2 (\wt y_4 - c_F(\lambda-\sgn(\lambda)))}{-q^0}\right). 
}
This result matches that obtained by performing the double Wick rotation against the T-channel transition matrix given by \eqref{TchannelTmatrix}.
Performing the duality transformation we obtain 
\be 
T_S = - T^{(B)}_S, 
\label{dualitySchannel}
\ee
where $T^{(B)}_S$ is given by \eqref{SchannelTmatrixboson}. 
% Thus the duality relation holds for S-channel scattering process. 

A main problem of the S-channel transition matrix determined in this way is that it does not satisfy unitarity condition \cite{Jain:2014nza}. 
In fact, the same problem of non-unitarity in S-matrix also exists in the original Aharonov-Bohm scattering \cite{Aharonov:1959fk}, which is to be obtained by taking the non-relativistic limit of the S-matrix in the current set up. 
The problem of non-unitarity in Aharonov-Bohm scattering was resolved by Ruijsenaars in \cite{Ruijsenaars:1981fp}, who pointed out that the delta-function type singular contribution in the forward scattering was missed in the original Aharonov-Bohm scattering and it becomes unitary by taking into account the singular contribution.
The reason why this phenomenon happens may be that only in the forward scattering the upper path of electron to the solenoid and lower one equally contribute to the wave function, and their coherent superposition gives the singular contribution. 

The method to cure the non-unitary problem in the relativistic situation  developed in \cite{Jain:2014nza} is also to take account of the singular contribution in the forward scattering by the ansatz of the Schwinger-Dyson equation of the scattering amplitude such that 
\be
T^{\rm conj}_S =T_1 (\sqrt s) + z T_2 (\sqrt s) + T_3(\sqrt s) {2\pi} \delta(\theta)
\label{ansatz}
\ee
where $\sqrt s = -q^0$ and $\theta$ is the scattering angle of an out-particle to the line formed by two in-particles in the center of mass frame \eqref{com}. 
By using $\theta$ the on-shell momenta of in and out particles can be parametrized as 
\be 
p^\mu=(E, p, 0), \quad k^\mu = (E, p\cos\theta, p\sin\theta), \quad E=\sqrt{p^2+c_F^2} = {\sqrt s \over 2} 
\label{SchannelMomenta}
\ee
which leads to $z = -i\cot{\theta\over2}$. 
$T_3$ describes the singular contribution in the forward scattering. 
Then the unitarity condition \eqref{unitaritySchannel} boils down to 
\beal{
-iT_1 + iT_1^*=&{1 \over 4 \sqrt s} (|T_1|^2 + T_1^* T_3 + T_1 T_3^* - |T_2|^2) ,
 \label{u1}\\
-iT_2 +iT_2^*=& {1 \over 4 \sqrt s} ( T_2^* T_3 + T_2 T_3^*),
 \label{u2}\\
-iT_3 + iT_3^*=&{1 \over 4 \sqrt s} ( |T_2|^2 + |T_3|^2),
\label{u3}
}
where we used $\sqrt s>2c_F$ in S-channel scattering. 

A solution physically preferable may be as follows. 
Since neither the relativistic nor non-relativistic situation seem to affect the argument for the singular term to be generated, 
it may be safe to assume that the singular part is the same as the non-relativistic case. 
\be
T_3={4i \sqrt s} (1- \cos \pi \lambda).
\label{T3} 
\ee 
Then \eqref{u2} is met if $T_2$ is real. 
In addition by using \eqref{u3} $T_2$ is determined as  
\be 
T_2={- 4 \sqrt s \sin \pi\lambda }. 
\label{T2}
\ee 
Finally \eqref{u1} is met if $T_3$ is given by 
\be 
T_1={ -4 \sqrt s } \sin (\pi\lambda) \tanh(X_F'(\sqrt s)).
\label{T1}
\ee

The solution constructed in this way possesses several desired properties as argued in \cite{Jain:2014nza} as follows. 
\begin{enumerate}[{(i)}]
 \item The unitarity is assured from construction. 

 \item This conjectural S-matrix in S-channel enjoys the bosonization duality. To see this let us compute the S-channel S-matrix by plugging the above solution into \eqref{SmatrixS}. The first term in \eqref{SmatrixS} by using \eqref{SchannelMomenta} can be computed as 
 \beal{ 
 {\bra{ -\vec r_4, -; -\vec r_1, +} \vec r_3, -; \vec r_2, +\rangle \over (2\pi)^3\delta^3(r_1+r_2+r_3+r_4)} =& { 8 \pi \sqrt s \delta(\theta) }. 
 }
 Then the result is 
 \beal{ 
 S_S^{\rm conj}= i k_F {\sin (\pi\lambda)\over \pi} T_S + {8\pi \sqrt s} \cos (\pi \lambda) \delta(\theta)
 \label{SchannelSmatrix}
 }
 where $T_S$ is given by \eqref{SchannelTmatrix}.%
 \footnote{ 
 The delta function $\delta(\theta)$ can be written in terms of the Mandelstam variables as $\delta(\theta) = \half \delta(\sqrt{t\over u})$. 
 }
 Due to \eqref{dualitySchannel} the conjectural S-matrix in S-channel transforms under the duality transformation as 
 \be
 S_S^{\rm conj}= -i k_B {\sin (\pi\lambda_B) \over \pi} T^{(B)}_S - {8\pi \sqrt s} \cos (\pi \lambda_B) \delta(\theta)= -S_S^{(B), \rm conj}, 
 \ee
 where in the second equation we used \eqref{SchannelSmatrixboson}.
 This shows that the physical scattering amplitudes coincide in the dual theories.
 
 \item Upon analytically continuing the S-channel S-matrix to off-shell region $\sqrt s < 2c_F$, a pole arises in a certain coupling region \cite{Dandekar:2014era}. The bound state energy is determined by 
 \be 
 y = \lambda +\frac{2 (\mf s+\sgn(\lambda)) \mf s}{(\mf s-\sgn(\lambda)) e^{2 \lambda  \tanh ^{-1}(\mf s)}+\mf s+\sgn(\lambda)}-\mf s-\sgn(\lambda)
 \ee
 where we set $\mf s = {\sqrt{s} \over 2c_F}, y={\wt y_4 \over c_F}$.  
 This indicates that a bound state of particle-antiparticle shows up when a solution exists: $1< {y \over \lambda} < {2-|\lambda| \over 1-|\lambda|}$.%
 \footnote{ 
 This region may be related to that where the theory possesses the reflection positivity. In fact, the reflection positivity of the two point correlator of the auxiliary field with fermion loops excluded implies that $k_F \wt y_4 >0$, which is compatible with the region for the pole to arise. 
 }
 In particular under the limit 
 $y_4, y_2^2 \to \infty$ with $\lambda, y_6, {y_2^2 \over y_4}$ fixed,
 in which this system reduces to the regular Chern-Simons fermionic vector model \cite{Minwalla:2015sca},
 the bound state disappears, which is consistent with the result in \cite{Frishman:2014cma}. 
 In addition the bound states in the dual theories map to each other under the duality transformation \cite{Inbasekar:2015tsa}.  
 
 \item The non-relativistic limit reduces the solution to the Aharonov-Bohm-Ruijsenaars scattering amplitude. To see this let us take the non-relativistic limit $\sqrt s \to 2c_B$ for the S-channel transition matrix with general coupling constants. The result is 
 \be 
 T^{\rm conj}_S \to 8c_F \(- \sin |\pi\lambda| -  \sin |\pi\lambda|i \cot {\theta' \over 2}  + i (1- \cos \pi \lambda) 2\pi\delta (\theta')\)
 \ee
 where we redefine the scattering angle so that  $\theta' = -\sgn(\lambda)\theta$. 
 In order to compare the scattering function in quantum mechanics, we scale the result by ${1\over c_F \sqrt p}$ (see \cite{Jain:2014nza}), which correctly reproduces the Aharonov-Bohm-Ruijsenaars scattering amplitude up to an overall numerical factor.  
 
 Connection to the self-adjoint extension of Aharonov-Bohm scattering \cite{AmelinoCamelia:1994we} was found in \cite{Dandekar:2014era}. This was achieved by taking the non-relativistic limit and simultaneously sending the couplings to the lower threshold for the bound state to exist, where the mass of the bound state approaches $2 c_F$. 
 By denoting deviations from the bounds by 
 \be
 \sqrt s = 2c_F + \epsilon_1 {p^2 \over c_F}, \quad 
 \wt y_4 = c_F(\lambda + \epsilon_2) 
 \ee
 the limit is given by 
 \be 
 \epsilon_1, \epsilon_2 \to 0, \quad {\epsilon_2 \over \epsilon_1^{\lambda_B}}: {\rm fixed} 
 \ee
 in the assumption that $\lambda<0$ or $\lambda_B>0$. 
 Under this limit the improved transition matrix in the bosonic side reduces to 
 \be 
  T^{\rm conj,(B)}_S \to - 8 c_B \[\sin \pi\lambda_B \( {\({2c_B \over p} \)^{2\lambda_B}{\epsilon_2 \over 2\epsilon_1^{\lambda_B}}e^{i\pi \lambda_B } -1 \over  {\({2c_B \over p} \)^{2\lambda_B} {\epsilon_2 \over 2\epsilon_1^{\lambda_B}}e^{i\pi \lambda_B } + 1 }} -i \cot {\theta \over 2} \) +i( \cos \pi \lambda_B -1) 2\pi\delta (\theta) \].
 \ee
 On the other hand, the Aharonov-Bohm scattering with a general self-adjoint boundary condition is given by \cite{AmelinoCamelia:1994we}
 \beal{
f(\theta) 
=& {-ie^{-i\pi \over 4} \over \sqrt{ 2\pi p} }\[ \sin \pi\alpha \( { { \Gamma(1+\alpha) \over w \Gamma(1-\alpha)} \({2 \over R p}\)^{2\alpha} e^{i\pi \alpha }-1 \over { \Gamma(1+\alpha) \over w \Gamma(1-\alpha)}\({2  \over R p}\)^{2\alpha} e^{i\pi \alpha }+1} - i \cot{\theta\over2} \)  + 2\pi i(\cos\alpha\pi -1)\delta(\theta) \]
 }
 where $p$ is the initial momentum, $\alpha$ is the background magnetic field normalized so as to be in the range from $0$ to $1$, and $R$ is the effective radius of the system. 
 They agree with each other up to an overall factor mentioned above under the identification such that 
 \be 
 {\epsilon_2 \over \epsilon_1^{\lambda_B}} = {2 \Gamma(1+\alpha) \over w \Gamma(1-\alpha)}, \quad c_B={1\over R}, \quad \lambda_B = \alpha.
 \label{identification}
 \ee
\end{enumerate}

Finally this result suggests that the usual crossing symmetry given in standard textbooks of quantum field theory needs to be modified when the scattering amplitude contains a singular contribution. 
In the current system upon performing analytic continuation from T-channel to S-channel the amplitude pick up the overall factor $N\times {\sin\pi\lambda \over \pi\lambda}$. 
The origin of this factor was discussed in relation to Wilson lines formed by trajectory of the charged particles \cite{Jain:2014nza}, though such a novel aspect is not confirmed from explicit calculation yet.  
We leave it to future works. 

\section{Discussion}
\label{discussion} 

We have computed the four point function of the fermionic field exactly in general $U(N)$ Chern-Simons fermionic vector model in the 't Hooft large $N$ limit. Applying the LSZ formula to this result with suitable Wick rotation we have calculated scattering amplitudes in T-channel and S-channel taking a special care for S-channel so as to include the singular contribution in the forward scattering to achieve unitarity. We have shown that the scattering amplitudes determined in this way enjoy the bosonization duality proposed in \cite{Minwalla:2015sca} as well as novel crossing relation argued in \cite{Jain:2014nza}. 
We have also shown that in the non-relativistic limit in a general coupling region the S-matrix reduces to the Aharonov-Bohm-Ruijsenaars scattering amplitude, and in a special coupling region of the threshold for the bound state to exist the S-matrix reduces to the self adjoint extension of Aharonov-Bohm scattering \cite{AmelinoCamelia:1994we} as claimed in \cite{Dandekar:2014era}.

It turned out that the conjectural S-matrix in S-channel in general Chern-Simons fermionic vector model develops a pole in a certain coupling region, which is a signal for a bound state of particle-antiparticle to exist. 
Analysis of a bound state in the regular Chern-Simons fermionic vector model was done from the Bethe-Salpeter equation with the result of no bound state in that system \cite{Frishman:2014cma}, which is consistent with the result in this paper.
It would be interesting to derive the Bethe-Salpeter equation in the current system and determine the bound state energy, which is expected to match that determined in this paper.  

As discussed in \cite{Minwalla:2015sca} 
the Chern-Simons vector models in the current paper are expected to be dual to a higher-spin gravity theory with parity violated by boundary condition  in AdS$_4$ \cite{Giombi:2011kc,Aharony:2011jz}. 
In the gravity side three point correlation functions were computed by taking advantage of infinitely many higher spin symmetries \cite{Giombi:2009wh,Giombi:2010vg,Giombi:2011ya}. (See also \cite{Giombi:2011rz,Maldacena:2012sf}.)  
It is interesting to compute the counterpart in the gravity side for the scattering amplitude computed in the paper. 
 
It would be interesting to recompute the scattering amplitude including chemical potential as done in \cite{Geracie:2015drf}, where the S-matrix in Chern-Simons theory with dense fermionic matter was computed and was applied to determining Landau parameters to see the consistency of microscopic calculation and thermodynamic macroscopic one.  
It would be intriguing to determine Landau parameters in the current setup and investigate its behavior under the RG flow and physical implication thereof. 

It would be also interesting to compute the scattering amplitude in higher supersymmetric case such as ABJ model by taking the vector model large $N$ limit. Some perturbative calculation of two body scattering in ABJ(M) theory was done in \cite{Agarwal:2008pu,Bianchi:2011fc,Chen:2011vv,Bianchi:2011dg,Bianchi:2012cq,Bargheer:2012cp,Bianchi:2014iia}. In the results there the singular term in the forward scattering does not show up, while the result in $\cN=1,2$ Chern-Simons vector model exhibits the singular contribution for the unitarity \cite{Inbasekar:2015tsa}. This issue may be cleared up by carrying out the large $N$ calculus developed in the analysis of Chern-Simons vector models explicitly. 

We hope to make progress in these issues in the near future. 

\section*{Acknowledgments}
The author would like to thank S. Jain, M. Mandlik and S. Minwalla for collaboration and valuable discussions. This work was supported by the MEXT-Supported Program for the Strategic Research Foundation at Private Universities “Topological Science” (Grant No. S1511006).

%%%%%%%%%%%%%%%%%%%%%%%%%%%%%%%%%%%%%%%%%%%%%%%%%%%%%%%%
\appendix 
\section{Regularization}
\label{regularization}

In this appendix we detail the regularization for the divergence we encountered in Section \ref{4pt} following \cite{Jain:2012qi}. 
The divergent integrals are contained in \eqref{B2condition}, \eqref{B4condition}, which are of a form such that 
\beal{
I=\int_{\mf h}^{\infty} {d h \over (2\pi)} 
{A_i(h) (2 i \Sigma_{I} + q_3 ) +2 B_i(h) (h^2 - \Sigma_{I}^2 ) \over 4 h^2 +q_3^2}
}
where $\mf h=\sqrt{\mf p_s^2 + c_F^2}$ and we denote $A_i(k_s',k_s)=A_i(h), B_i(k_s',k_s)=B_i(h)$ with $h=h_{k'}=\sqrt{(k_s')^2+c_F^2}$. 
Using \eqref{AtoB} and \eqref{Bsol} one can show that the asymptotic behavior of the numerator in the integrand around $h\sim\infty$ is 
\beal{
A_i(h) (2 i \Sigma_{I} + q_3 ) +2 B_i(h) (h^2 - \Sigma_{I}^2 )
= 2 B_i(\infty) h^2 + \cO(h^0).
}
Using this we can divide the above divergent integral into two parts in a way that 
\beal{
I=\int_{\mf h}^{\infty} {d h \over (2\pi)} \[
{A_i(h) (2 i \Sigma_{I} + q_3 ) +2 B_i(h) (h^2 - \Sigma_{I}^2 ) -  2 B_i(\infty) h^2  \over 4 h^2 +q_3^2}
+ {2 B_i(\infty) h^2  \over 4 h^2 +q_3^2}\].
}
The first term, which we denote by $I_1$, is convergent and computed as 
\beal{
I_1=&\lim_{\Lambda\to\infty} \int_{\mf h}^{\Lambda} {d h \over (2\pi)} \[
{A_i(h) (2 i \Sigma_{I} + q_3 ) +2 B_i(h) (h^2 - \Sigma_{I}^2 ) -  2 B_i(\Lambda) h^2  \over 4 h^2 +q_3^2}\] \nn
=&\frac{1}{16 \pi  \lambda }\bigg\{\alpha_i \lambda  \left(\pi  \sgn(q_3)-2 \tan ^{-1}\left(\frac{2 \mf h}{q_3}\right)\right)+2 \beta_i f(\mf p_s) (2 c_F (\lambda-\sgn(\lambda))-2 \mf h \lambda +i q_3) \nn
&+\beta_i f(\infty ) \left(-4 c_F (\lambda-\sgn(\lambda))+4 \mf h \lambda -2 \lambda  q_3 \tan ^{-1}\left(\frac{2 \mf h}{q_3}\right)+\pi  \lambda  q_3 \sgn(q_3)-2 i q_3\right)\bigg\}.
}
The second term, which we denote by $I_2$, is a divergent term, which we can easily regularize by the dimensional regularization prescribed in \cite{Jain:2012qi}. 
For this purpose we rewrite the integral as the original three-dimensional integral form.  
\be 
I_2 =2 B_i(\infty) \int {d^3 k' \over (2\pi)^3} { (k_s')^2 + c_F^2  \over (k'^2+c_F^2)((k'+q)^2+ c_F^2)}{k'_-\over (k' - \mf p)_-}.
\ee
Then we can apply the dimensional regularization prescribed in \cite{Jain:2012qi} to this integration.
\be 
I_2 = 2 B_i(\infty) \int {d^2 k' d^{1-\epsilon} \mbf k_3 \over (2\pi)^3} { h^2  \over ((\mbf k_3)^2 + h^2)((\mbf k_3+ \mbf q_3)^2 + h^2)}{k'_-\over (k' - \mf p)_-}.
\ee
This can be computed in a standard method by employing Feynman parameters. The result is 
\be 
I_2 = 2 B_i(\infty) \times {-1 \over 8\pi}(\mf h - {q_3 \over 2} \tan^{-1}{2 \mf h \over q_3} + {\pi |q_3| \over 4}). 
\ee
Finally we obtain the regulated integral as 
\beal{
I=&I_1+I_2 \nn 
=&\frac{-2 \alpha_i \mf h \lambda +\beta_i q_3 f(\mf p_s) (2 c_F (\lambda-\sgn(\lambda))-2 \mf h \lambda +i q_3)+\beta_i q_3 f(\infty ) (-2 c_F \lambda +2 c_F-i q_3)}{8 \pi  \lambda  q_3}.
}

\section{Construction of asymptotic states}
\label{states}

In this appendix we construct asymptotic particle states which are necessary to define the S-matrix. 
For this purpose we quantize the fermionic field of the system \eqref{csfnonlinear2} in the canonical formulation.  
The Dirac equation of the system \eqref{csfnonlinear2} is 
\be 
(\gamma_\mu \partial^\mu + \Sigma_F) \psi^m_x =0 
\ee
where $m$ is the fundamental index of the U($N$) gauge group and $\Sigma_F$ is given by \eqref{1piselfenergy}. 
To find a complete set of the solutions we do Fourier transformation: 
\be
( i \gamma_\mu p^\mu + \Sigma_F(p)) \psi^m_p =0. \quad 
% \bar\psi_{-p}( i \gamma_\mu p^\mu +\Sigma_F(p))  =0. 
\label{Diraceq} 
\ee
A non-trivial solution of the Dirac equation exists only when the momentum is on mass-shell: 
\be
p^0=\pm E_{\vec p}, \quad E_{\vec p}:= \sqrt{\vec p^2 +c_{F}^2}.
\ee
Then a positive-energy solution is found to be 
\be 
u({\vec p})
=  (\frac{ -ip_3+\Sigma_I(p)}{\sqrt{p^1+p^0}}, -i\sqrt{p^1+p^0}), \quad 
\label{u}
\ee
with $p^0=E_{\vec p}$,  
and negative-energy one is 
\be
v({-\vec p})
=(\frac{i p_3-\Sigma_I(p)}{\sqrt{p^1+p^0}}, i{\sqrt{p^1+p^0}}), \quad
\label{v} 
\ee
with $p^0=-E_{\vec p}$. 
These are normalized in such a way that 
\beal{
\bar u({\vec p}) u({\vec p}) = 2  \Sigma_I(p), \quad 
\bar v({-\vec p}) v({-\vec p}) = - 2 \Sigma_I(p). \quad 
}
Then we expand the fermionic field in terms of this complete set. 
The result is 
\beal{
\psi^m_x  
=&  \int {d^2 p \over (2\pi)^2} {1 \over \sqrt{2E_{\vec p}}}  \(a^m_{\vec p} u({\vec p}) e^{i x p} +(b_{m,\vec p})^\dagger v({\vec p}) e^{- i x p} \)|_{p^0=E_{\vec p}}
}
where $a_{\vec p}, b_{\vec p}$ are expansion coefficients of the field. 

Following the canonical formalism we introduce the canonical commutation relation. 
\beal{
\{ a^m_{\vec p}, (a^n_{\vec q})^\dagger \} = (2\pi)^2 \delta^m_n \delta^2(\vec p - \vec q), 
\quad 
\{ b_{m,\vec p}, (b_{n,\vec q})^\dagger \} = (2\pi)^2 \delta_m^n \delta^2(\vec p - \vec q). 
}
Then $(a^m_{\vec p})^\dagger, (b_{m,\vec p})^\dagger$ are one-particle creation operators with the positive, negative charges, respectively. 
We call the particle with positive charge antiparticle. 
By using these creation operators we can define in-state and out-state of one particle by 
\beal{
&\ket{\vec p,+,m} =\sqrt{2E_{\vec p}} (a^m_{\vec p})^\dagger \ket {0},\quad 
\ket{\vec p,-,m} =\sqrt{2E_{\vec p}} (b_{m,\vec p})^\dagger \ket {0},\quad \nn
&\bra{\vec p,+,m} = \sqrt{2E_{\vec p}} \bra{0} a^m_{\vec p} ,\quad 
\bra{\vec p,-,m} =\sqrt{2E_{\vec p}} \bra{0} b_{m,\vec p},\quad 
}
where $\ket{0}$ is the vacuum state. 
Here the normalization is determined so that the inner product of these state becomes Lorentz invariant:
\beal{
&\bra{\vec p, +,m} \vec q, +,n \rangle = \bra{\vec p, -,n} \vec q, -,m \rangle
 = 2E_{\vec p}  (2\pi)^2 \delta^m_n \delta^2(\vec q - \vec p). 
}
In the main text we study two particle scattering, where two particle states are defined by 
\bes{
&\ket{\vec q,-,n;\vec p,+,m} =\sqrt{2E_{\vec q}} (b_{n,\vec q})^\dagger \sqrt{2E_{\vec p}} (a^m_{\vec p})^\dagger\ket {0},\\
&\bra{\vec q,-,n;\vec p,+,m} = \sqrt{2E_{\vec p}}\sqrt{2E_{\vec q}} \bra{0}a^m_{\vec p}b_{n,\vec q}. 
\label{2particle}
}

\section{S-matrix in general Chern-Simons bosonic vector theory} 
\label{bosonicSmatrix}

In this appendix we give a brief derivation of two body scattering amplitude in general Chern-Simons vector model in the 't Hooft large $N$ limit. 
This can be done by applying the method developed in \cite{Jain:2014nza} to the current case, in which the triple trace coupling is included. 
Lagrangian of general Chern-Simons bosonic vector model is \cite{Minwalla:2015sca,Jain:2013gza} 
\beal{
\cL_B=&i \varepsilon^{\mu\nu\rho}{k_B \over 4 \pi}
\Tr( A_\mu\partial_\nu A_\rho -{2 i\over3}  A_\mu A_\nu A_\rho)
+ D_\mu \bar \phi D^\mu\phi   \nn
&+m_B^2 \bar\phi \phi +  {4\pi b_4 \over k_B}(\bar\phi \phi)^2
+ {(2\pi)^2 x_6 \over (k_B)^2} (\bar\phi \phi)^3.
\label{cssaction}
}

The exact propagator of the scalar field is determined as \cite{Jain:2013gza}
\beal{
\langle \phi^m(p) {\bar \phi}_n(-p') \rangle &=\delta^m_n (2 \pi)^3  \delta^3(-p'+p) \alpha_B(p), \quad 
\alpha_B(p) = {1 \over p^2+ c_B^2}
}
where $c_B^2$ is the physical mass of the scalar field determined by the following gap equation 
\beal{
c_{B}^2 =& \lambda_B^2 (1+3x_6) {c_{B}^2 \over 4} -2\lambda_B  b_4 c_{B}+ m_B^2. 
}

As in the fermionic case in order to compute the four point correlator we determine the four point vertex given by 
\be 
-\half \int{d^3 p\over(2\pi)^3}{d^3 k\over(2\pi)^3}{d^3 q\over(2\pi)^3} \bar\phi_{m}(-p-q) \phi^{m}(p) B(p, k; q) \bar\phi_{n}(-k) \phi^{n}(k+q). 
\ee
As the fermionic case, the bootstrap diagram for this four point vertex is precisely the same as the ladder diagram in the 't Hooft large $N$ limit, thus the Schwinger-Dyson equation is 
\beal{
&B(p, k; q) = B^0 (p, k; q) + N \int {d^3 k' \over (2\pi)^3} B^0 (p, k'; q) \alpha_B(q+k') \alpha_B(k') B(k', k; q). 
\label{bx40}
}
Here $B^0$ represents the contribution of the ladder bar, which is diagrammatically given by Figure 4 in \cite{Jain:2014nza} with the contribution of the triple trace vertex given by  
\beal{ 
- \( { 2\pi \over k_B } \)^2 \times 3! \times N \int {d^3 r \over (2\pi)^3} \alpha_B(r) 
=  6 x_6 \( { 2\pi \over k_B } \)^2 N { c_{B}\over 4\pi}. 
\label{phi6} 
}
Thus we obtain  
\beal{
B^0 (p, k; q) =& { 4\pi \over ik_B } q_3 z + {{\mf b}_4\over k_B}
\label{b0}
}
where $\mf b_4$ is a constant given by  
\be
{\mf b}_4= { 2\pi \lambda_B} c_{B}- {8\pi b_4 } + { 6 \pi x_6 \lambda_B } c_{B}. 
\ee

The bootstrap equation \eqref{bx40} can be solved exactly in the frame \eqref{com}. 
In order to solve \eqref{bx40} we set the ansatz \cite{Aharony:2012nh} 
\be
B (p, k; q) = B_1 (p_s, k_s) + q_3 z  B_2 (p_s, k_s), 
\label{bansatz}
\ee
where $z$ is defined by \eqref{z}.
Plugging this into \eqref{bx40} and performing $k_3$-integral and angular integral for $k_-$ we find
\beal{
B_2 (p_s, k_s)=& {4\pi \over ik_B} + N \int_{h_p}^{h_k} {dh \over (2\pi)} 
{1 \over \( 4 h^2 + q_3^2\)} {- 8\pi q_3 \over ik_B}  B_2,  
\label{b2}\\
B_1 (p_s, k_s)=& {{\mf b}_4\over k_B} + 
N \int^{h_p}_{h_0} {dh  \over (2\pi)} {1 \over \( 4h^2 + q_3^2\)} {8\pi q_3 \over ik_B}  B_1 +N \int_{h_0}^{h_k} {dh \over (2\pi)} {1 \over \( 4h^2 + q_3^2\)} {- 2 {\mf b}_4q_3 \over k_B}  B_2 \nn 
&+N \int_{h_0}^{\infty} {dh \over (2\pi)} {1 \over \( 4h^2 + q_3^2\)} \( {- 4\pi q_3 \over ik_B} + { {\mf b}_4\over k_B}\) \biggl(B_1+ q_3  B_2 \biggl), 
\label{b1}
}
where in the right-hand side we set $B_i=B_i(k_s',k_s)$ with $h=h_{k'}$. 
A solution of these integral equations is 
\beal{ 
B_2 (p_s, k_s) = & {4 \pi \over ik_B} { f_B(p_s) \over f_B(k_s)}, \\
B_1 (p_s, k_s) = &{4 \pi q_3 \over ik_B}  
\[- { ( {\mf b}_4+i 4 \pi q_3)f_B(\infty) + ( {\mf b}_4-i 4 \pi q_3) f_B(0) \over  ( {\mf b}_4+i 4 \pi q_3)f_B(\infty) - ({\mf b}_4- i 4 \pi q_3) f_B(0)} \] {f_B(p_s) \over f_B(k_s)}, 
\label{Bsol}
}
where $f_B(p_s)$ is defined by 
\be
f_B(p_s) := \exp\[ -2i\lambda_B \arctan {2 \sqrt {p_s^2 +c_B^2} \over q_3 }  \]. 
\label{fb}
\ee
Therefore 
\beal{
B (p, k; q)=& {4 \pi q_3 \over i k_B}  ( z + i \tan X_B(q_3) ) {f_B(p_s) \over f_B(k_s)} 
\label{Bsol}
}
where $X_B(q_3)$ is given by \eqref{XB}. 

Then the procedure to compute scattering amplitude is completely parallel to the fermionic case described in the main text. 
Therefore we do not repeat the description for the bosonic case, and we present only results with the subscript $(B)$ for the corresponding quantities. 
The T-channel transition matrix is 
\beal{
T_T^{(B)} =  {4 \pi q_3 \over i k_B}  ( z + i \tan X_B(q_3) ) 
=-\frac{4 \pi}{k_B} \(\sigma\sqrt{su \over -t} - \sqrt{-s}\tan X_B(\sqrt{-s}) \), 
\label{TchannelTmatrixboson}
}
and the naive S-channel transition matrix obtained by double Wick rotation is 
\be
T_S^{(B)} =  {4 \pi q^0 \over k_B}  ( z + \tanh X_B'(-q^0) ) 
=-\frac{4 \pi}{k_B} \(\sigma\sqrt{su \over -t} - \sqrt{s}\tanh X_B'(\sqrt{s}) \), 
\label{SchannelTmatrixboson}
\ee
where $s,t,u$ variables are defined by \eqref{stu} and 
\be 
X_B'(\sqrt s) = \lambda_B \tanh^{-1}{\sqrt s \over 2c_B} + \tanh^{-1}\({-4b_4 +c_B \lambda_B (1+ 3x_6) \over 2 \sqrt s} \). 
\ee
By making the S-channel S-matrix unitary as done in \cite{Jain:2014nza} we obtain 
\be
S_S^{(B), \rm conj}= i k_B {\sin (\pi\lambda_B) \over \pi} T^{(B)}_S + {8\pi \sqrt s} \cos (\pi \lambda_B) \delta(\theta). 
\label{SchannelSmatrixboson}
\ee

%%%%%%%%%%%%%%%%%%%%%%%%%%%%%%%%%%%%%%%%%%%%%%%%%%%%%%%%
\bibliographystyle{utphys}
\bibliography{CSGN}
\end{document}